\documentclass[prd,superscriptaddress,amsfonts,amssymb,amsmath,showpacs,twocolumn]{revtex4-2}

\usepackage{bm}
\usepackage{amsfonts}
\usepackage{latexsym}
\usepackage[latin1]{inputenc}
\usepackage{graphicx}
\usepackage{amsmath}
\usepackage{palatino}
\usepackage{mathpazo}
\usepackage{textcomp}
 \usepackage{cuted}
\usepackage{float}
\usepackage{booktabs}
\usepackage{dcolumn}
\usepackage{ragged2e}
\usepackage{hyperref}
\hypersetup{colorlinks,citecolor=blue}
\hypersetup{colorlinks=true,linkcolor=red,filecolor=magenta,    urlcolor=blue}
\usepackage{amsmath}
\usepackage{graphicx}
\usepackage{makecell}
\usepackage{hyperref}
\usepackage[mathlines]{lineno}
\usepackage{amssymb}
\usepackage{enumitem}
\usepackage{mathtools}
\usepackage{mathrsfs}
\usepackage{setspace}
\usepackage{color}
\usepackage{graphicx}
\usepackage[dvipsnames]{xcolor}
\usepackage{mathalfa}
\usepackage{calligra}
\usepackage[misc]{ifsym}
\DeclareMathAlphabet{\mathpzc}{OT1}{pzc}{m}{it}
\DeclareMathAlphabet{\mathcalligra}{T1}{calligra}{m}{n}
\hypersetup{colorlinks=true,citecolor=blue,linkcolor=red,urlcolor=magenta}
\usepackage{xcolor}
\usepackage{orcidlink}
\usepackage[caption=false]{subfig}
\usepackage{commath}
\usepackage{multirow}
\def\jnl@style{}
\def\aaref@jnl#1{{\jnl@style#1}}

\def\aaref@jnl#1{{\jnl@style#1}}

\def\aj{\aaref@jnl{AJ}}                   
\def\apj{\aaref@jnl{ApJ}}                 
\def\apjl{\aaref@jnl{ApJ}}                
\def\apjs{\aaref@jnl{ApJS}}               
\def\apss{\aaref@jnl{Ap\&SS}}             
\def\aap{\aaref@jnl{A\&A}}                
\def\aapr{\aaref@jnl{A\&A~Rev.}}          
\def\aaps{\aaref@jnl{A\&AS}}              
\def\mnras{\aaref@jnl{Mon.~Not.~Roy.~Astron.~Soc.}}             
\def\prd{\aaref@jnl{Phys.~Rev.~D}}        
\def\plb{\aaref@jnl{Phys.~Lett.~B}}        
\def\prc{\aaref@jnl{Phys.~Rev.~C}}  
\def\prl{\aaref@jnl{Phys.~Rev.~Lett.}}    
\def\qjras{\aaref@jnl{QJRAS}}             
\def\skytel{\aaref@jnl{S\&T}}             
\def\ssr{\aaref@jnl{Space~Sci.~Rev.}}     
\def\zap{\aaref@jnl{ZAp}}                 
\def\nat{\aaref@jnl{Nature}}              
\def\aplett{\aaref@jnl{Astrophys.~Lett.}} 
\def\apspr{\aaref@jnl{Astrophys.~Space~Phys.~Res.}} 
\def\physrep{\aaref@jnl{Phys.~Rep.}}      
\def\physscr{\aaref@jnl{Phys.~Scr}}       
\def\commat{\aaref@jnl{Comm.~Math.~Phys.}}              
\def\science{\aaref@jnl{Science}}               
\def\cqg{\aaref@jnl{Classical Quant.~Grav.}}            
\def\jpcs{\aaref@jnl{JPCS}}                                     
\def\ijmpd{\aaref@jnl{Int.~J.~Mod.~Phys.~D}}                    
\def\grg{\aaref@jnl{Gen.~Relat.~Gravit.}}               
\def\rpp{\aaref@jnl{Rep.~Prog.~Phys.}}          
\def\npa{\aaref@jnl{Nucl.~Phys.~A}}        
\def\lrr{\aaref@jnl{Living Rev.~Rel.}}                   
\def\jcap{\aaref@jnl{J.~Cosmology Astropart.~Phys.}}    
\def\rmp{\aaref@jnl{Rev.~Mod.~Phys.}}   
\def\epjc{\aaref@jnl{Eur.~Phys.~J.~C}}

\allowdisplaybreaks[1]

\begin{document}

\preprint{APS/123-QED}

\title{Charged Gravastars with Casimir Energy in Quadratic f(T) Gravity}


 \author{S. R. Harshitha\orcidlink{0009-0003-1942-6327 }}
\email{harshitha8.sr@gmail.com}
\affiliation{Department of P.G. Studies and Research in Mathematics,
 \\
 Kuvempu University, Shankaraghatta, Shivamogga 577451, Karnataka, INDIA
}%

\author{C. S. Varsha\orcidlink{0009-0008-3529-0390}}
\email{varshaagowda0902@gmail.com}
\affiliation{Department of P.G. Studies and Research in Mathematics,
 \\
 Kuvempu University, Shankaraghatta, Shivamogga 577451, Karnataka, INDIA
}
\affiliation{Department of Sciences and Humanities, CHRIST (Deemed to be University) School of Engineering and
Technology,\\
Kanmanike, Kumbalgodu, Mysore Road, Bangalore-560074, Karnataka, India,
}%

\author{V. Venkatesha\orcidlink{0000-0002-2799-2535}}%
 \email{vensmath@gmail.com}
\affiliation{Department of P.G. Studies and Research in Mathematics,
 \\
 Kuvempu University, Shankaraghatta, Shivamogga 577451, Karnataka, INDIA
}%

\author{Phalaksha Murthy B}\orcidlink{0000-0001-7125-6749}%
\email{pmurthymath@gmail.com} \affiliation{Government First Grade College Kadur, Karnataka, 577548, India}

\date{\today}

\begin{abstract}
 We investigate a charged gravastar configuration in the framework of modified teleparallel gravity, considering the quadratic form $f(T)=T+\alpha T^{2}$ and incorporating Casimir vacuum energy as a source for the interior region. The gravastar is constructed as a three-region configuration consisting of a Casimir-energy dominated vacuum interior, an intermediate thin shell of ultra-relativistic matter satisfying the equation of state $p=\rho$, and an exterior Reissner-Nordstrom spacetime. The modified field equations are derived for a static, spherically symmetric geometry, and the interior and shell solutions are obtained by incorporating the contributions of Casimir energy and the electromagnetic field. The matching between the interior and exterior geometries is performed using the Israel-Darmois junction conditions, from which the surface energy density and pressure of the shell are determined. We examine the physical properties of the configuration through the proper length, shell energy, and entropy. The influence of the charge parameter on these quantities is also investigated. The resulting profiles remain finite and smooth across the shell, while the proper length, energy, and entropy exhibit regular behaviour with increasing shell thickness. The model provides a framework for studying the interplay between quantum vacuum effects, electric charge, and torsional modifications of gravity in compact-object configurations.

\textbf{keywords: Charged garavastar, Casimir energy, $f(T)$ gravity, Reissener-Nordstrom metric.} 
\end{abstract}

\maketitle

\tableofcontents
 \section{Introduction}
The final fate of gravitational collapse remains one of the most profound open problems in theoretical physics. Within classical General Relativity, the collapse of a sufficiently massive star inevitably leads to the formation of a black hole \citep{Penrose:1964wq,Penrose:1969pc,Oppenheimer:1939ue,Hawking:1976ra,Hawking:1975vcx} characterized by an event horizon and a central singularity where curvature diverges and known laws of physics break down. These features give rise to conceptual difficulties, including the information loss paradox and the problem of singularities, motivating the search for alternative compact object models that remain regular everywhere\citep{Carr:1974nx, Preskill:1992tc,Chen:2014jwq}.

An important proposal in this direction is the \textbf{gravastar (gravitational vacuum condensate star)}, originally introduced by \textbf
{Mazur and Mottola}\citep{Mazur:2001fv}. In this picture, the classical black hole interior is replaced by a region of vacuum energy with negative pressure, matched to an exterior Schwarzschild spacetime through a thin shell of ultra-relativistic matter. The gravastar thus avoids horizon formation and central singularities while remaining observationally similar to black holes in many astrophysical contexts. The physical motivation for such a structure arises naturally from considering quantum vacuum effects in strong gravitational fields.

Vacuum fluctuations and the Casimir effect \citep{DeWitt:1975ys,Bordag:2001qi,Casimir:1948dh,Klimchitskaya:2009cw}provide direct evidence that quantum fields possess measurable zero-point energy. Although the Casimir force has been extensively verified in laboratory experiments, its gravitational implications are still under active investigation. Mazur and Mottola\citep{Mazur:2004ku} emphasized that vacuum energy need not be determined by microscopic cutoffs alone, but may instead arise from \textbf{boundary conditions} in the large, analogous to the Casimir effect in finite cavities. They proposed that a gravitational vacuum phase transition near the classical horizon could generate an interior region of effective vacuum energy, supporting a non-singular compact object configuration.

In this framework, the gravastar interior behaves as a de Sitter like vacuum, sustained by negative pressure, while the thin shell corresponds to a quantum transition layer separating regions with different vacuum energy densities. Such a configuration naturally regularizes the divergences of the stress-energy tensor near classical horizons and provides a macroscopic realization of a gravitational Casimir like boundary effect. This perspective establishes a deep connection between semiclassical gravity, vacuum fluctuations, and compact astrophysical objects.

More recently, attention has turned toward refinement of gravastar models by including additional physical ingredients such as electric charge\citep{Usmani:2010ac,Yousaf:2019zcb,Sharif:2021sjm,Debnath:2019eor}, anisotropic pressures\citep{Cattoen:2005he,Maurya:2017uyk,Beltracchi:2024dfb}, and modified theories of gravity\citep{Das:2017rhi,Shamir:2018qhq,Bhatti:2020cjz,Ghosh:2020rau,Yousaf:2020nza,Sengupta:2020lhw}. The inclusion of \textbf{electric charge} is particularly relevant, as it modifies the exterior spacetime to a Reissner-Nordstrom geometry and introduces repulsive Coulomb forces that can further stabilize compact configurations. Charged gravastars \citep{Bhattacharjee:2024xjf, Mustafa:2024xjs, Usmani:2010ac, Rahaman:2012wc, Ghosh:2015ohi} thus provide a richer phenomenology and allow investigation of the interplay between electromagnetic fields, vacuum polarization, and gravitational structure.

Motivated by these developments, the present work investigates a \textbf{charged gravastar model supported by Casimir energy}. Instead of assuming an \textit{ad hoc} cosmological constant in the interior, we consider vacuum energy arising from quantum fluctuations subject to gravitational boundary conditions, analogous to the Casimir effect. This approach provides a physically motivated source of negative pressure and offers a natural mechanism for sustaining the gravastar core. The thin shell is modeled as a transitional layer where the vacuum structure rearranges, while the exterior geometry is taken to be that of a charged compact object.

The study of Casimir supported gravastars is particularly appealing because it links observable quantum vacuum phenomena to astrophysical scale gravitational systems. Moreover, incorporating electric charge allows exploration of stability domains, surface redshift bounds, and mass-radius relations that differ from classical black holes and uncharged gravastars \citep{Chavanis:2019bnu,Heggie:2006se,Chirenti:2007mk}. Such models may offer testable signatures in strong field astrophysical observations and gravitational wave data.

This paper is organized as follows. \autoref{section2} presents the basic formalism of $f(T)$ gravity and derives the corresponding field equations for a static spherically symmetric spacetime. In \autoref{section3}, the coupling of $f(T)$ gravity with Casimir energy is developed, incorporating vacuum fluctuations and charge contributions into the effective energy-momentum tensor. The detailed gravastar configuration is constructed in \autoref{section4}, where the interior region, thin shell, and exterior spacetime are obtained separately. \autoref{section5} is devoted to the Israel-Darmois junction conditions, which determine the surface stresses and ensure smooth matching across the shell. The physical characteristics of the model, including proper length, energy content, and entropy, are analyzed in \autoref{section6}. \autoref{section7} discuss the stability of the configuration through entropy maximisation. Finally, in \autoref{section8} we provide the conclusion of our analysis. 

\section{$f(T)$ gravity}\label{section2}
General Relativity can be reformulated in a \textit{teleparallel} framework by taking the field of orthonormal frames. In $f(T)$ gravity, the gravitational field is described by a set of four linearly independent tetrad fields $h_a(x^\mu)$ where $a= 0,1,2,3.$ These tetrads define an orthonormal frame in the tangent space at each spacetime, satisfying,

\begin{equation}
    h_a. h_b = \eta _{ab},
\end{equation}

with $\eta_{ab} = diag (1,-1,-1,-1)$ the Minkowski metric. Expressed in a coordinate basis, each tetrad vector takes the form $h_a= h^\mu _a \partial_\mu.$ Here, Latin indices label tangent-space components, whereas Greek indices denote spacetime coordinates. The spacetime metric arises naturally from the tetrads through,
\begin{equation}
    g_{\mu \nu(x)}= \eta _{ab} h^a_{\mu(x)} h^b_{\nu(x)},
\end{equation}

where $\mu$ and $\nu$ are the coordinate indices of the manifold, and also pass over $0,1,2,3$, and $h^\mu _a$ forms the tangent vector of the manifold.

In General Relativity, gravitation is described using the torsion-free Levi-Civita connection. In contrast, teleparallel gravity formulates gravitational interactions through the Weitzenb\"ock connection, which is curvature-free but possesses non-vanishing torsion. The Weitzenb\"ock connection is defined by,
\begin{equation}
T^ \lambda _{\mu \nu} = h^\lambda _a(\partial_\nu h^a_\mu+ w^a_{b\nu} h^b_\nu),
\end{equation}
where $w^a_{b\mu}$ denotes the spin connection associated with inertial effects. This spin connection vanishes in a suitable inertial frame where gravitational effects are locally removed. It may be determined through the reference tetrad $h^a_{(r)\mu}$ corresponding to the gravity-free configuration as,
\begin{equation}
w^a_{b\mu} = \frac{1}{2} h_{(r)c\mu} (f^{bac}_{(r)}+f^{cab}_{(r)}-f^{abc}_{(r)}),
\end{equation}
 where the structure coefficients $f^c_{ab}$ are obtained from the tetrad through,
\begin{equation}
f^c_{ab} = h^\mu _a h^\nu_b(\partial_\nu h^c_\mu - \partial_\mu h^c_\nu).
\end{equation}
Using the tetrad and spin connection, the torsion tensor takes the form,
\begin{equation}
T^\lambda_{\mu\nu} = h^\lambda _a(\partial_\mu h^a_\nu -\partial_\nu h^a_\mu+w^a_{b\mu} h^b_\nu-w^a_{b\nu} h^b_\mu).
\end{equation}
From the torsion tensor, one introduces the contortion tensor,
\begin{equation}
K^{\mu \nu}{}_\rho =- 1/2 (T^{\mu\nu}{}_\rho-T^{\nu \mu}{}_\rho -T_\rho{}^{\mu \nu}).
\end{equation}
and the superpotential tensor, 
\begin{equation}
S_\rho{}^{\mu\nu} = 1/2 (K^{\mu \nu}{}_\rho+ \partial^\mu_\rho T^{\alpha \nu}{}_\alpha - \partial^\nu_\rho T^{\alpha \mu}{}_\alpha).
\end{equation}
The torsion scalar is then constructed as
\begin{equation}
T=S_\rho {}^{\mu \nu}T^{\rho}{}_{\mu \nu}.
\end{equation} 
When the torsion scalar $T$ is used as the gravitational Lagrangian, the resulting theory is the Teleparallel Equivalent of General Relativity (TEGR), as $T$ and $R$ differ only by a total divergence term. Extending this framework, $f(T)$ gravity is obtained by promoting the Lagrangian to a general function of $T$, resulting in the action
\begin{equation}
    S=\int{d^4x\frac{h}{16\pi G}f(T)},
\end{equation}
where, $ h = det(h^a_\mu) = \sqrt{-g}. $
Varying this action with respect to the tetrad yields the vacuum field equations of $ f(T) $gravity as,
\begin{widetext}
\begin{equation}\label{field equ}
h^{-1}\partial_\nu (h  S_a{}^{\mu \nu})f_T+S_a{}^{\mu \nu}\partial_\nu (T)f_{TT} - h^\lambda _a T^\rho{} _{\nu \lambda} S_\rho{}^{\mu \nu} f_T+1/4 h^\mu _a f(T)= 4\pi G h^\rho _a T_\rho{}^\mu,
\end{equation}    
\end{widetext}
where, $S_a{}^{\mu\nu}= h_a^\rho S_\rho{}^{\mu\nu}$ $f_T=df/{dT}$ and $f_{TT}= d^2f/{dT^2}$ and $T^\mu_\rho$ is the energy-momentum tensor of matter.

Contracting with $h^a_\nu$, the equations may be expressed in spacetime index form as,  
\begin{widetext}
\begin{equation}
\partial _\sigma (h S_\nu{}^{\mu \sigma})f_T + h S_\nu{}^{\mu \sigma}(\partial_\sigma T)f_{TT}-h T^\rho{}_{\sigma \nu} S_\rho{}^{\mu \sigma} f_T+h/4 \partial^\mu_\nu f(T) = 4\pi G h T^\mu _\nu.
\end{equation}
\end{widetext}

\subsection*{Specialization to static spherical symmetry}

We consider a static and spherically symmetric spacetime configuration. This assumption implies that the system exhibits neither rotational nor translational dynamics. Accordingly, the spacetime geometry may be represented by the line element,
\begin{equation}
ds^2= e^{\nu(r)}dt^2-e^{\lambda(r)}dr^2-r^2 (d \theta ^2+ sin^2 \theta d\phi^2),
\end{equation}
where the metric functions $\nu(r)$ and $\lambda(r)$ depend solely on the radial coordinate $r$.
To preserve invariance of the line element under local Lorentz transformations, we adopt a diagonal tetrad basis of the form
\begin{equation}
[e^i_\mu]=\begin{pmatrix}
   e^{\nu(r)/2} & 0 & 0 & 0\\
    0 & e^{\lambda(r)/2}& 0 & 0\\
    0 & 0 & r & 0\\
    0 & 0 & 0 & rsin\theta
\end{pmatrix}.
\end{equation}
The determinant of the tetrad matrix then follows as
\begin{equation}
    e= det[e^i_\mu]= e^{(\nu+\lambda)/2}r^2 sin\theta.
\end{equation}
Using this tetrad choice, the torsion scalar associated with the Weitzenb\"ock connection becomes,
\begin{equation}
    T(r)= \frac{2e^{-\lambda}}{r} \left(\nu'(r)+\frac{1}{r}\right),
\end{equation}
and a suitable diagonal tetrad choice, the field equations yield three independent components.
Inserting the components $S^{\mu\nu}_a$ and $T^a_{\mu\nu}$ in Equation \eqref{field equ} one can obtain the field equations as, 
\begin{widetext}
\begin{equation}
  4\pi\rho= \frac{-e^{-\lambda(r)}}{r} T'(r)f_{TT}-\left[\frac{T(r)}{2}-\frac{1}{2r^2}-\frac{e^{-\lambda(r)}}{2r}(\nu'(r)+\lambda'(r))\right]f_T+\frac{f}{4} ,
 \end{equation}
 \end{widetext}
 \begin{widetext}
    \begin{equation}
    4\pi p_r=\left[\frac{T(r)}{2}-\frac{1}{2r^2}\right]f_{TT}-\frac{f}{4},
      \end{equation}
       \end{widetext}
        \begin{widetext}
      \begin{equation}
    4\pi p_t =-\frac{e^{-\lambda(r)}}{2}\left(\frac{\nu'(r)}{2}+\frac{1}{r}\right)T'(r) f_{TT}+\frac{T(r)}{4}f_T+\frac{e^{-\lambda(r)}}{2}\left[  \frac{\nu''(r)}{2}+\left( \frac{\nu'(r)}{4}+\frac{1}{2r}\right) (\nu'(r)-\lambda'(r))\right]f_T-\frac{f}{4},
    \end{equation}
    \end{widetext}
    and
    \begin{equation}
    \frac{e^{-\lambda(r)/2} Cot\theta}{2r^2}T'(r)f_{TT}=0.
    \end{equation}

where prime(') denotes the derivatives with respect to $r$ and $\rho$, $p_r$, $p_t$ represent the energy density, radial pressure and tangential pressure, respectively.
By taking the covariant differentiation on both sides of Equation \eqref{field equ}, one gets the non-conservation law of the energy-momentum tensor,
\begin{equation}
    \frac{dp}{dr}+\nu'(p+\rho)=-\frac{1}{8\pi r^2}f_{TT}\frac{dT}{dr}.
\label{eq:conservation}
\end{equation}

In the present work, we adopt the quadratic power-law model of (f(T)) gravity given by,
\begin{equation}
    f(T)=T+\alpha T^n.
\end{equation} 
where $T$ denotes the torsion scalar and $\alpha$ is a constant parameter characterizing the deviation from the teleparallel equivalent of General Relativity (TEGR). The first term represents the TEGR contribution, while the quadratic term introduces nonlinear torsion corrections that become significant in strong gravitational regimes.
The derivatives of the chosen model are,
$f_T= 1+n \alpha T^{n-1} $ and $f_{TT}=n (n-1) \alpha T ^{n-2} $
For this paper, taking $n=2$ we get,
\begin{equation}
    f(T)=T+\alpha T^2.
\end{equation} 
$f_T= \frac{df}{dT}= 1+2 \alpha T $ and $f_{TT}= \frac{d^2f}{dT^2}=2\alpha $.
After substituting the quadratic power-law model into the modified field equations, the resulting field equations are obtained as follows:

\begin{widetext}
\begin{equation}
    4\pi \rho = \frac{1}{4}(T(r)+\alpha T^2(r))-\frac{2 e^{-\lambda }\alpha T'(r)}{r}-(1+2 \alpha T(r))\left(  - \frac{1}{2r^2}+\frac{T(r)}{2}-\frac{e^{-\lambda(r)}(\lambda'(r)+\nu'(r))}{2r}\right),
\end{equation}
\begin{equation}
    4\pi p_r = \frac{1}{4} (-T(r)-\alpha T^2(r))+\left( - \frac{1}{2r^2}+\frac{T(r)}{2}\right)(1+2 \alpha T(r)),
\end{equation}
\end{widetext}
and
\begin{widetext}
\begin{equation}
\begin{split}
   & 4\pi p_t = \frac{1}{4}(-T(r)-\alpha T^2(r))+ \frac{1}{4} T(r)(1+2 \alpha T(r))+e^{-\lambda(r)} \alpha T'(r)\left(\frac{1}{r}+\frac{\nu'(r)}{2}\right)+\frac{1}{2} e^{-\lambda(r)}(1+2 \alpha T(r))\\
   &\left( \left( \frac{1}{2r}+\frac{\nu'(r)}{4}\right) (-\lambda'(r) + \nu'(r)+\frac{\nu''(r)}{2}\right). \\
    \end{split}
\end{equation}
\end{widetext}
\section{Coupling of $f(T)$ gravity with Casimir energy}\label{section3}
The \textit{Casimir effect} \citep{DeWitt:1975ys,Bordag:2001qi,Casimir:1948dh,Klimchitskaya:2009cw} arises between two perfectly conducting, uncharged plane-parallel plates placed at a very small separation in vacuum. It was first predicted theoretically by Casimir in 1948\citep{Budkov:2024hey} and later confirmed experimentally at the Philips laboratories. However, in recent years, high-precision experiments have provided more reliable verification of this phenomenon.

The \textit{Casimir energy} is a change in the vacuum energy of a quantum field caused by boundary conditions. It is not energy added by matter or particles but an energy that already exists in the vacuum and becomes measurable when boundaries are imposed.

In quantum field theory, the vacuum state is
not trivially empty but is characterised by persistent zero-point
fluctuations of all quantum fields. Each normal mode of a
quantised field of frequency $\omega$ contributes a zero-point
energy $E_{\rm ZPE} = \frac{1}{2}\hbar\omega$, so the total
vacuum energy is formally $E_{\rm vac} = \sum_{k}\frac{1}{2}\hbar\omega_{k}$.
Although this sum diverges, only differences in vacuum
energy are physically observable. When boundary conditions are
imposed. Such as perfectly conducting plates or the
spherical shell bounding the gravastar core, the spectrum
of allowed vacuum modes is altered, producing a finite and
measurable shift in zero-point energy:
\begin{equation}
    \Delta E= E_{cas}= E_{vac} (\text{with boundaries}) - E_{vac}(\text{free space}).
\end{equation}
This finite energy difference is called the Casimir energy.
\\

For two perfectly conducting parallel plates separated by distance $a$, the Casimir effect gives rise to a negative vacuum energy density take the well-known form \citep{Garattini:2019ivd},
\begin{equation}
    \rho_c=-\frac{hc\pi^2}{720a^4},
\end{equation}
which is \emph{negative}, confirming that the vacuum energy
inside the cavity is lower than in free space. The associated
Casimir pressure follows from the thermodynamic relation
$P = -dE/dV$, giving
\begin{equation}
    P_c= -3 \rho_c = \frac{hc\pi^2}{240 a^4}.
\end{equation}
\subsection{Casimir Energy as the Interior Source}
Mazur and Mottola \cite{Mazur:2001fv} emphasised that vacuum energy
need not be fixed solely by microscopic ultraviolet cutoffs; it
may arise instead from macroscopic boundary conditions analogous
to the Casimir effect in finite cavities. In the gravastar
picture, a gravitational vacuum phase transition near the would-be
horizon generates an interior region of effective vacuum energy.
The thin shell separating the interior from the exterior
corresponds to a quantum transition layer across which the vacuum
energy density changes abruptly, playing the role of the
conducting plates in the gravitational Casimir analogy. This
perspective motivates the use of equation
as the source of negative pressure in the gravastar interior,
rather than an ad hoc cosmological constant.

In the present model we adopt a spherically symmetric
generalisation of casimir density equation by replacing the
plate separation $a$ with the radial coordinate $r$, following
the prescription of \citep{Garattini:2019ivd}. The resulting Casimir
energy density and pressure in curved spacetime are
\begin{equation}
    \rho_c(r)= - \frac{hc\pi^2}{720 r^4} , \quad
      p_c^{(r)}(r)= p_c^{(t)}(r) =+\frac{hc\pi^2}{240 r^4}
\end{equation}
\subsection{Electromagnetic Energy-momentum Tensor}
For a static spherically symmetric charged compact object, the
only non-vanishing component of the Faraday tensor is the radial
electric field $F_{01} = -F_{10} = Q/r^{2}$, where $Q$ is the
total charge. The electromagnetic energy-momentum tensor
$T^{\mu\nu}_{\rm EM} = \frac{1}{4\pi}
(F^{\mu\alpha}F^{\nu}_{\ \alpha} - \frac{1}{4}g^{\mu\nu}
F_{\alpha\beta}F^{\alpha\beta})$ yields mixed components,
\begin{equation}
    \rho_{EM}\;=\; T^0_{0 (EM)} \;=\; +\frac{Q}{8 \pi r^4},
\end{equation}
\begin{equation}
    p_{r (EM)}\;=\; -T^1_{1 (EM)} \;=\; -\frac{Q}{8 \pi r^4},
\end{equation}
and
\begin{equation}
     p_{t (EM)}\;=\; -T^2_{2 (EM)} \;=\; +\frac{Q}{8 \pi r^4},
\end{equation}
The radial and tangential electromagnetic pressures carry
\emph{opposite signs}, reflecting the tensile nature of the
Maxwell stress tensor in the radial direction. This anisotropy
is a purely electromagnetic effect, independent of the modified
gravity framework, and is consistent with the result of
Cattoen, Faber, and Visser \cite{Cattoen2005} that gravastars
must possess anisotropic pressures. A direct consequence is the
non-vanishing anisotropy parameter
\begin{equation}
    \Pi \;\equiv\; p_{t} - p_{r}
        \;=\; \frac{Q^{2}}{4\pi r^{4}},
    \label{eq:anisotropy}
\end{equation}
which increases toward the interior of the shell and vanishes
in the uncharged limit $Q \to 0$.
\\
In the present model the total energy-momentum tensor on the
right-hand side of the field equations receives contributions from three independent
sources: ordinary matter (the gravastar fluid), Casimir vacuum
energy, and the electromagnetic field. These combine into
effective quantities that differ between the energy density
and the two pressure components:

\begin{table}[h]
\centering
\caption{Contributions to the effective energy-momentum
         tensor from matter, Casimir, and electromagnetic
         sources. The sign of the electromagnetic term
         reverses between radial and tangential pressure,
         generating pressure anisotropy.}
         \vspace{0.3 cm}
\label{tab:sources}
\begin{tabular}{lccc}
\hline\hline
Source
  & $\rho_{r,{\rm eff}}$
  & $p_{r,{\rm eff}}$
  & $p_{t,{\rm eff}}$ \\
\hline
Matter
  & $+\rho$
  & $+p$
  & $+p$ \\
Casimir
  & $-\dfrac{\hbar c\pi^{2}}{720r^{4}}$
  & $+\dfrac{\hbar c\pi^{2}}{240r^{4}}$
  & $+\dfrac{\hbar c\pi^{2}}{240r^{4}}$ \\[6pt]
Electromagnetic
  & $+\dfrac{Q^{2}}{8\pi r^{4}}$
  & $-\dfrac{Q^{2}}{8\pi r^{4}}$
  & $+\dfrac{Q^{2}}{8\pi r^{4}}$ \\
\hline\hline
\end{tabular}
\end{table}
The modified field equations governing the three distinct regions of the gravastar become,
\begin{widetext}
\begin{equation}
    4\pi \left(\rho-\frac{hc\pi^2}{240 r^4}+\frac{Q^2}{8 \pi r^4}\right) = \frac{1}{4}(T(r)+\alpha T^2(r))-\frac{2 e^{-\lambda }\alpha T'(r)}{r}-(1+2 \alpha T(r))\left(  - \frac{1}{2r^2}+\frac{T(r)}{2}-\frac{e^{-\lambda(r)}(\lambda'(r)+\nu'(r))}{2r}\right),
     \label{eq:modified_Eq1}
\end{equation}
\begin{equation}
    4\pi \left(p + \frac{hc\pi^2}{240 r^4}-\frac{Q^2}{8 \pi r^4}\right) = \frac{1}{4} (-T(r)-\alpha T^2(r))+\left( - \frac{1}{2r^2}+\frac{T(r)}{2}\right)(1+2 \alpha T(r)),
    \label{eq:modified_Eq2}
\end{equation}
\end{widetext}
and
\begin{widetext}
\begin{equation}
\begin{split}
   & 4\pi \left( p+\frac{hc \pi^2}{240 r^4}+ \frac{Q^2}{8 \pi r^4 }\right) = \frac{1}{4}(-T(r)-\alpha T^2(r))+ \frac{1}{4} T(r)(1+2 \alpha T(r))+e^{-\lambda(r)} \alpha T'(r)\left(\frac{1}{r}+\frac{\nu'(r)}{2}\right)+\frac{1}{2} e^{-\lambda(r)}(1+2 \alpha T(r))\\
   &\left( \left( \frac{1}{2r}+\frac{\nu'(r)}{4}\right) (-\lambda'(r) + \nu'(r)+\frac{\nu''(r)}{2}\right). \\
     \label{eq:modified_Eq3}
    \end{split}
\end{equation}
\end{widetext}
\section{Gravastar model}\label{section4}
 In the gravastar paradigm, gravitational collapse \citep{Hawking:1975vcx} does not culminate in the formation of a singularity or an event horizon. Instead, a quantum phase transition near the classical
Schwarzschild radius replaces the black hole interior with a
region of vacuum energy, producing a compact configuration
composed of three distinct regions: an interior
vacuum core dominated by Casimir energy, an intermediate thin
shell of ultra-relativistic stiff matter, and an exterior vacuum
spacetime. The line element for the static, spherically symmetric
spacetime is\citep{Mazur:2001fv}.
\begin{equation}
    ds^{2} = e^{\nu(r)}\,dt^{2}
           - e^{\lambda(r)}\,dr^{2}
           - r^{2}(d\theta^{2} + \sin^{2}\theta\,d\varphi^{2}),
    \label{eq:metric}
\end{equation}
where the metric functions $\nu(r)$ and $\lambda(r)$ are
determined separately in each region and matched across the
shell boundaries.

 \subsection{Interior region}
  \subsubsection{Physical ansatz and equation of state}
  The innermost region $0 \leq r \leq D$ is dominated by Casimir
vacuum energy and is modelled as a de Sitter-like condensate
following the original proposal of Mazur and Mottola \cite{Mazur:2001fv}.
We adopt the equation of state
\begin{equation}
    p \;=\; -\rho, \qquad \rho \;=\; \rho_{0} = \mathrm{const},
    \label{eq:interior_EOS}
\end{equation}
where $\rho_{0}$ is a constant energy density sourced primarily
by the Casimir vacuum. The condition $\rho = \mathrm{const}$ is
an ansatz consistent with a homogeneous de Sitter core; it is
not independently derived from the field equations but is
physically motivated by the Mazur-Mottola picture of a
gravitational vacuum condensate. This equation of state
corresponds to the dark energy parameter $\omega = -1$ and
generates a repulsive gravitational effect that prevents the
formation of a central singularity.
\subsubsection{Torsion scalar is constant in the interior}
Substituting \eqref{eq:interior_EOS} into the modified
conservation equation \eqref{eq:conservation}, the left-hand
side vanishes identically: $dp/dr = 0$ since $\rho_{0}$ is
constant. The conservation equation then reduces to
\begin{equation}
    0 \;=\; -\frac{\alpha}{4\pi r^{2}}\frac{dT}{dr},
    \label{eq:interior_conservation}
\end{equation}
which, for $\alpha \neq 0$, forces
\begin{equation}
\begin{aligned}
    &\frac{dT}{dr} \;=\; 0
    \quad\Longrightarrow\quad
    T(r) \;=\; T_{0} \;=\; \mathrm{const}\\
   & \qquad \text{throughout the interior}.
    \label{eq:T_constant}
    \end{aligned}
\end{equation}
The constancy of the
torsion scalar in the interior follows necessarily from the de
Sitter ansatz and the nonlinear $f(T)$ conservation law. We note
that in the linear case $\alpha = 0$ this argument does not apply;
the interior torsion is then constrained by the field equations
alone. The effective $f(T)$ coupling factor evaluated at $T = T_{0}$
is
\begin{equation}
    \mathcal{F} \;\equiv\; 1 + 2\alpha T_{0},
    \label{eq:F_def}
\end{equation}
which is a genuine constant in the interior. In the GR limit
$\alpha \to 0$, $\mathcal{F} \to 1$ and the standard
teleparallel result is recovered.
\subsubsection{Field equations in the interior}
 
With $T = T_{0}$ constant, $T'(r) = 0$, and $\mathcal{F}$
a constant, the three field equations
\eqref{eq:modified_Eq1}--\eqref{eq:modified_Eq3} simplify.
The $f_{TT}T'$ term in Eq.~\eqref{eq:modified_Eq1} vanishes,
and adding Eqs.~\eqref{eq:modified_Eq1} and
\eqref{eq:modified_Eq2} with $\rho_{\rm eff} + p_{\rm eff}$
computed from Table~\ref{tab:sources} gives

\begin{equation}
    4\pi(\rho_{\rm eff} + p_{\rm eff})
    \;=\; \frac{\mathcal{F}\,e^{-\lambda}}{2r}(\nu' + \lambda').
    \label{eq:sum_interior}
\end{equation}
Evaluating the left-hand side: the matter terms cancel
($\rho_{0} - \rho_{0} = 0$), the electromagnetic terms cancel
($\rho_{\rm EM} + p_{r,\rm EM} = 0$), and only the Casimir
contribution survives:
\begin{equation}
    \rho_{\rm eff} + p_{\rm eff}
    \;=\; -\frac{\hbar c\pi^{2}}{720 r^{4}}
         + \frac{\hbar c\pi^{2}}{240 r^{4}}
    \;=\; \frac{\hbar c\pi^{2}}{360 r^{4}}.
    \label{eq:rho_plus_p_interior}
\end{equation}
This gives the relation
\begin{equation}
    \nu'(r) + \lambda'(r)
    \;=\; \frac{\hbar c\pi^{3}}{45\,\mathcal{F}\,r^{3}\,e^{-\lambda(r)}}.
    \label{eq:nu_lam_sum}
\end{equation}

We eliminate $\nu'$ using the torsion constraint. From the
definition $T_{0} = 2e^{-\lambda}(\nu' + 1/r)/r$, with
$u(r) \equiv e^{-\lambda(r)}$:
\begin{equation}
    \nu'(r) \;=\; \frac{T_{0}\,r}{2\,u(r)} - \frac{1}{r},
    \qquad
    \lambda'(r) \;=\; -\frac{u'(r)}{u(r)}.
    \label{eq:torsion_sub}
\end{equation}
a constant, we simplify the three field equations
\eqref{eq:modified_Eq1}--\eqref{eq:modified_Eq3}. Now, from  considering \eqref{eq:modified_Eq1} and \eqref{eq:interior_EOS} we find the metric potential $\lambda$ as,
ubstituting \eqref{eq:torsion_sub} into
Eq.~\eqref{eq:modified_Eq1} and collecting terms in $u$ and
$u'$, the energy density equation reduces to the
linear first-order ordinary differential equation
\begin{equation}
    u'(r) + \frac{u(r)}{r}
    \;=\; \frac{1}{r}
         - \frac{\alpha T_{0}^{2}\,r}{2\mathcal{F}}
         - \frac{8\pi\rho_{0}\,r}{\mathcal{F}}
         + \frac{\hbar c\pi^{2}}{180\,\mathcal{F}\,r^{3}}
         - \frac{Q^{2}}{2\,\mathcal{F}\,r^{3}}.
    \label{eq:interior_ODE}
\end{equation}
 Integrating term by term and imposing the regularity
condition $C_{1} = 0$ to remove the $C_{1}/r$ singularity at
$r = 0$, we obtain,
\begin{equation}
      e^{-\lambda}(r)
    \;=\; 1
         - \frac{\alpha T_{0}^{2}\,r^{2}}{6\mathcal{F}}
         - \frac{8\pi\rho_{0}\,r^{2}}{3\mathcal{F}}
         - \frac{\hbar c\pi^{3}}{90\,\mathcal{F}\,r^{2}}
         + \frac{Q^{2}}{\mathcal{F}\,r^{2}}
           \label{eq:interior_metric}
        \end{equation}
This result has been verified by direct substitution into
Eq.~\eqref{eq:interior_ODE}, confirming it satisfies the ODE
identically. In the GR limit $\alpha \to 0$, $\mathcal{F} \to 1$,
and it reduces to the standard
charged gravastar interior metric. Each term in Eq.~\eqref{eq:interior_metric} 
has a clear physical origin: the
constant $1$ corresponds to flat space, the $\rho_{0}$ term
reflects the de Sitter energy density, the $\alpha T_{0}^{2}$
term is the genuine $f(T)$ nonlinear correction, the
$\hbar c\pi^{2}$ term encodes Casimir energy, and the $Q^{2}$
term represents the electromagnetic contribution.

With $e^{-\lambda}(r)$ known from \eqref{eq:interior_metric},
$\nu'(r)$ is obtained directly from the torsion constraint
\eqref{eq:torsion_sub}:
\begin{equation}
    \nu'(r) \;=\; \frac{T_{0}\,r}{2\,e^{-\lambda}(r)} - \frac{1}{r}.
    \label{eq:nu_prime_interior}
\end{equation}
Integrating \eqref{eq:nu_prime_interior} gives $\nu(r)$ up to
an integration constant $c_{3}$, which is fixed by requiring
continuity of $\nu$ at the inner shell boundary $r = D$. Since
the integrand involves a rational function of $r$ with
denominator $e^{-\lambda}(r)$, the integral is evaluated
numerically using the explicit expression \eqref{eq:interior_metric}. 

The active gravitational mass of the interior region is,
\begin{equation}
    M(D)= \int_{0}^{r_1 =D} 4\pi r^2 \rho_0 dr= 4/3 \pi D^3\rho_0.
\end{equation}
 \subsection{Thin shell}
 Surrounding the vacuum core is a thin shell $D\leq r \leq D+\epsilon$ composed of ultra-relativistic matter obeying the stiff equation of state,
\begin{equation}
    p=\rho.
\end{equation}
This shell represents the compressed remnant of the original collapsing star and serves as a boundary layer connecting the interior de Sitter region to the exterior vacuum spacetime. The surface energy density and surface pressure on the shell are determined through the Israel junction conditions, which ensure the continuity of the spacetime metric across the interfaces. The thin shell maintains mechanical equilibrium between the outward repulsive force generated by the interior vacuum pressure and the inward gravitational attraction of the exterior region.
Within the non-vacuum region, namely the shell, obtaining exact solutions of the field equations is generally quite challenging. However, with the thin-shell approximation, characterized by $0<e^{-\lambda}\ll1$ and $\epsilon \ll 1$, it becomes possible to derive approximate analytical solutions. Following the prescription of Israel, the intermediate region connects the two spacetime manifolds. In this region, physical quantities that depend on the radial coordinate 
$r$ vary only slightly and may be approximated as nearly constant across the shell.
\subsubsection{Key structural difference from the interior}
Unlike the interior, the torsion scalar $T(r)$ is not constant in the shell. With $p=\rho$, the conservation equation does not reduce to $T'=0$
Instead it becomes,
\begin{equation}
    \frac{d\rho}{dr}+\nu'(r) \rho = -\frac{\alpha}{4 \pi r^2}\frac{dT}{dr},
\end{equation}
which is a genuine first-order linear ODE coupling $\rho(r)$ and $T(r)$. The varying $T(r)$ means $f_{TT} T' = 2 \alpha T' \neq 0$, so the non-linear $f(T)$ correction actively contributes to all three field equations in the shell. This is the key structural distinction between the interior (where nonlinearity enters only through $\mathcal{F}$) and the shell (where $T'$ drives and additional dynamical source). 
Subtracting Eq2 from Eq 1 with $p=\rho$ and applying the thin shell approximation, gives the governing equation for $e^{-\lambda(r)}$ in the shell.
\begin{equation}
    \frac{Q^2}{r^4}-\frac{hc\pi^2}{45 r^4} \;=\; -\frac{1}{2r}\frac{d}{dr}(e^{-\lambda(r)})+\frac{1}{r^2}
    \label{eq:shell ODE}
\end{equation}
This equation is purely geometric. The matter density $\rho$ has been eliminated and reflects the balance between the electromagnetic and Casimir source terms and the geometry of the shell.
On integrating Eq \eqref{eq:shell ODE} we get,
\begin{equation}
    e^{-\lambda(r)}\;=\; -\frac{hc\pi^2}{45 r^2}+\frac{Q^2}{r^2}+2 \log r +c_2
  \label{eq: thin shell}
    \end{equation}
where $c_2$ is an integration constant.
With $\nu'(r) \approx \nu'_D$ constant (Israel approximation), the lapse function in the shell is simply
\begin{equation}
    \nu(r) \;=\; \nu'_Dr +c_3
\end{equation}
where, $c_3$ is fixed by continuity of $\nu$ at $r=D$. The value of $\nu'(D)$ is obtained from the torsion constraint evaluated at $r=D$;
\begin{equation}
    \nu'_D = \frac{T_D D}{2 e^{-\lambda(D)}}-\frac{1}{D}
\end{equation}
where $T_D=T_0$, continuity of the torsion scalar at the inner boundary.
\subsubsection{Shell density}
With the metric functions $e^{-\lambda}(r)$ and $\nu(r)$
determined in the thin-shell approximation, the energy density
$\rho(r)$ of the stiff matter constituting the shell is obtained
from the energy density field equation \eqref{eq:modified_Eq1}.
Substituting the shell metric \eqref{eq: thin shell}, the
frozen lapse derivative $\nu'(r) \approx \nu_D'$, and the
torsion scalar profile into the left-hand side of
Eq.~\eqref{eq:modified_Eq1}, and collecting terms in $\rho$
and $\rho'$, solving the equation we get,
\begin{widetext}
\begin{equation}
\begin{split}
    \rho(r) = & e^{-r \nu}c_1 + \frac{1}{270} e^{-r \nu} \left( \frac{1}{r^4} e^{r \nu} \big( 540 r^2(1+3 r \nu)+c h \pi^2(6+ 10 r \nu+ 5 r^2 \nu^2+5 r^3 \nu^3)-45 Q^2(6+10 r \nu+5 r^2 \nu^2+5 r^3 \nu^3) \right)\\
    & \frac{1}{270} e^{-r \nu}(-5 \nu^2(324 + ch \pi^2 \nu^2-45 Q^2 \nu^2) \operatorname{Ei}(r \nu) ).
 \end{split} 
 \label{eq: p=r}
\end{equation}
\end{widetext}

\begin{figure}[!ht]
\centering
    \includegraphics[width=1\linewidth]{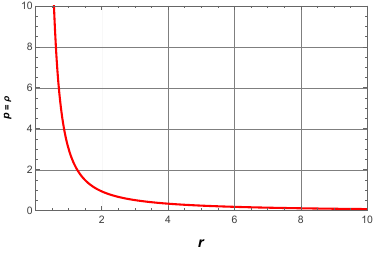}
    \caption{ {Variation of the energy density $\rho$ (equal to pressure 
$p$) with respect to the radial coordinate $r$ in the shell region. The coincident profiles confirm the stiff fluid equation of state $p=\rho$
 and indicate a smooth and well-behaved matter distribution throughout the shell.} }
    \label{fig2}
\end{figure}

The plot above (\autoref{fig2}) illustrates the radial behavior of the energy density and the pressure $p=\rho$. The profiles are positive and monotonically decreasing with r. A relatively rapid decrease is observed at smaller radii, whereas the profiles become progressively flatter as r increases, tending towards negligible values at larger radii. The smooth and positive behaviour of these quantities indicates a physically well- behaved matter distribution within the region considered.
\subsection{Exterior region}
The spacetime outside the shell is vacuum and is described by the Schwarzschild solution (or its appropriate generalization in modified gravity theories) \citep{Pradhan:2023xyr,Pradhan:2023wac}. Consequently, the external gravitational field of a gravastar is identical to that of a black hole of the same mass, rendering the object observationally indistinguishable from a classical black hole at large distances. However, unlike black holes, gravastars possess neither an event horizon nor a central singularity.
The exterior region obeys the EOS,
\begin{equation}
    p=\rho=0.
\end{equation} 
The line element for the exterior region is the Reissener- Nordstrom metric, which is given by,
\begin{widetext}
    \begin{equation}
    ds^2=-\left( 1-\frac{2M}{r}+\frac{Q^2}{r^2}\right)dt^2+\left( 1-\frac{2M}{r}+\frac{Q^2}{r^2}\right)^{-1}dr^2+r^2(d\theta^2 +sin^2\theta d\phi^2).
\end{equation}
\end{widetext}
\section{Junction condition}\label{section5}
In gravastar constructions, the spacetime is separated into an interior vacuum region and an exterior vacuum region, which are joined across a timelike hypersurface representing a thin shell of matter. Since interior and exterior geometries are described by different metric solutions, a consistent matching across the shell is required. This matching is carried out using the Israel-Darmois junction formalism \citep{Israel:1966rt,Senovilla:2013vra}.

The first junction condition requires the continuity of the induced metric across the hypersurface, ensuring that the spacetime geometry is regular at the shell and that no intrinsic metric singularity is present. This condition guarantees that physical quantities such as proper time and spatial distances remain well defined for observers crossing the shell.

The second junction condition allows for a discontinuity in the extrinsic curvature across the hypersurface. Such a discontinuity is interpreted as the presence of matter confined to the thin shell. The jump in extrinsic curvature determines the surface stress-energy tensor from which the surface energy density and surface pressure of the shell are obtained.

In the gravastar framework, the interior vacuum region generates a repulsive gravitational effect, while the exterior spacetime produces an attractive gravitational field. The thin shell provides the necessary surface stresses to balance these opposing effects and stabilize the configuration. The Israel-Darmois formalism thus plays a crucial role in determining the physical properties of the shell and ensuring the overall consistency of the gravastar model.

The hypersurface $\Sigma$ represents the three-dimensional thin-shell, while $\kappa^+$ and $\kappa^-$ denote the exterior and interior regions, respectively. Thus, the complete spacetime manifold is given by $\kappa^+ \cup \Sigma \cup \kappa^- $. The physical characteristics of the shell are described in terms of the surface energy density $\sigma$ and the surface pressure $p$.
The interior spacetime is described by the static spherically symmetric metric
\begin{equation}
ds^2= -e^{\nu(r)}dt^2+e^{\lambda(r)}dr^2+r^2 (d \theta ^2+ sin^2 \theta d\phi^2),
\end{equation}
To describe the exterior region, we consider a general static spherically symmetric metric of the form 
\begin{equation}
    ds^2=-F(r) dt^2 +F(r)^{-1}dr^2 +r^2 (d\theta^2+sin^2 \theta d\phi^2).
\end{equation}
According to Birkhoff's theorem, the exterior geometry of a static, spherically symmetric gravastar is uniquely determined and is given by the Reissner-Nordstrom spacetime. It is worth noting that the spatial sections in both the interior and exterior regions preserve spherical symmetry.

On the junction hypersurface, the induced metric reduces to the Friedmann-Lemaitre-Robertson-Walker (FLRW) form,
\begin{equation}
    ds^2 = -d\tau^2 +a(\tau)d \Omega^2.
\end{equation}
where, $\tau$ denotes the proper time on the shell, $a(\tau)$, $a(tau)$ is the shell's radius and $d\Omega^2$ represents the metric on the unit two-sphere.

Since the interior and exterior geometries differ, the extrinsic curvature of the junction hypersurface must be computed separately on each side, leading to a possible discontinuity that encodes the physical properties of the thin shell. This is given by
\begin{equation}
   K^{\pm}_{ij} = -n^{\pm}_\mu \left(\frac{\partial^2 x^\mu}{\partial \xi^i \partial \xi^j}
+ T^\mu_{\alpha\beta} \frac{\partial x^\mu}{\partial \xi^i} \frac{\partial x^\beta}{\partial \xi^j}\right),
\end{equation}
where $\xi$ denotes the intrinsic coordinates defined on the shell, and $n^{\pm}_\mu$ are the unit normal vectors to the junction hypersurface evaluated from the interior $(-)$ and exterior $(+)$ regions.
\begin{equation}
    n^{\pm}_\mu=\pm \left| g^{\mu \beta}\frac{\partial f}{\partial x^\mu}\frac{\partial f}{\partial x^\beta}\right|^{-1/2} \frac{\partial f}{\partial x^\mu}
\end{equation}
where $n_\mu$ is a unit timelike four-vector satisfying the normalization condition 
$n_\mu n^\mu=1$.The surface energy density and pressure required to maintain the stability of the thin shell are then determined using the Lanczos equations.
Within this framework, the intrinsic surface stress-energy tensor $S^i_j$ on the thin shell is governed by \textbf{Lanczos equations}, which are expressed as,
\begin{equation}
    S^i_j= -\frac{1}{8\pi}(k^i_j-\partial^i_jk^k_k ).
\end{equation}
where $k^i_j$ denotes the discontinuity in the second fundamental form across the hypersurface and is defined by, 
\begin{equation}
    k_{ij}= K^+_{ij}-K^-_{ij}
\end{equation}
where the indices $i,j=0,2,3$ correspond to the intrinsic coordinates of the shell. Since the radial coordinate $r$ remains constant at the junction surface, the surface stress-energy tensor can be written diagonally as,
\begin{equation}
    S_{ij}= diag(-\varsigma,p),
\end{equation}
where $\varsigma$ and $p$ are the surface energy density and the surface pressure at the junction surface $r=D$, respectively, expressed as:
\begin{equation}\label{eq52}
    \varsigma=-\frac{1}{4\pi D}[\sqrt{f}]^+_-
\end{equation}
and
\begin{equation}\label{eq53}
    p=-\frac{\varsigma}{2} +\frac{1}{16\pi}\left[\frac{f'}{\sqrt{f}}\right]^+_-
\end{equation}
and also Equation-of-state (EoS)
\begin{equation}
    \omega=\frac{p}{\varsigma}.
\end{equation} 
Using Equations \eqref{eq52} and \eqref{eq53}, we obtain 
\begin{widetext}

\begin{equation}
    \varsigma= -\frac{1}{4D\pi}\left( \sqrt{1-\frac{2M}{D}+\frac{q^2}{D^2}}-\sqrt{1-\frac{\alpha T_0^2 D^2}{6 F}-\frac{ch\pi^3}{90 F D^2}-\frac{8 \pi \rho_0 D ^2}{6 F}-\frac{Q^2}{F D^2} }\right),
\end{equation}
\begin{equation}
\begin{split}
    p=&\frac{1}{8 D \pi}\left(\frac{1}{\sqrt{\frac{D^2-2 D M+ q^2}{D^2}}}+ \frac{\sqrt{10} D^2(\alpha T_0^2+16 \pi \rho_0)}{F \sqrt{\frac{90 D^2 F- c h \pi^3+ 90 Q^2-15 D^4(\alpha T_0^2+16 \pi \rho_0)}{D^2 F}}}\right)\\
    &\frac{1}{8 D \pi}\left(-\frac{M}{D \sqrt{\frac{D^2-2 D M +q^2}{D^2}}}-\frac{3 \sqrt{10}}{\sqrt{\frac{90 D^2 F - c h \pi^3 + 90 Q^2 - 15 D^4 (\alpha T_0^2-16 \pi \rho_0)}{D^2 F}}}\right)\\
    \end{split}
\end{equation}

\begin{equation}
   \omega=-\frac{1}{2}-\frac{\frac{30(D M -q^2)}{\sqrt{\frac{D^2-2 D M+q^2}{D^2}}}+ \frac{\sqrt{10}(-c h \pi^3 +90 Q^2 +15 D^4 \alpha T_0^2+240 D^4 \pi \rho_0)}{F\sqrt{\frac{90 D^2 F - c h \pi^3+ 90 Q ^2 -15 D^4 \alpha T_0^2- 240 D^4 \pi \rho_0}{D^2 F}}}}{60 D^2 \left( \sqrt{\frac{D^2- 2 D M +q^2}{D^2}}- \frac{\sqrt{\frac{90 D^2 F - c h \pi^3 +90 Q^2 -15 D^4 \alpha T_0^2 -240 D^4 \pi \rho_0}{D^2 F}}}{3 \sqrt{10}}\right)}
\end{equation}  
    
\end{widetext}

A set of well-established criteria, known as the \textbf{energy conditions}, must be satisfied for a geometric configuration to be physically acceptable. The commonly considered energy conditions are the null, weak, strong, and dominant energy conditions, which are expressed as follows:
\begin{itemize}
    \item {Null Energy Condition(NEC):  $\varsigma+p>0$,}
    \item {Weak Energy Condition(WEC):  $\varsigma>0,\varsigma+p>0$,}
    \item {Strong Energy Condition(SEC):  $\varsigma+p>0, \varsigma+3p>0$,}
    \item {Dominant Energy Condition(DEC):  $\varsigma>0, \varsigma\pm p>0$}.
\end{itemize}
The physical viability of the proposed gravastar model is ensured if these energy conditions are satisfied. In the present analysis, we focus on the \textbf{null energy condition}, which plays a central role in determining whether the thin shell is composed of ordinary or exotic matter. It is important to note that a violation of the NEC automatically implies a violation of the remaining energy conditions.

Furthermore, the variation in surface energy density with respect to the shell thickness parameter $(\epsilon)$ is clearly shown in the figure, clearly indicating that the surface energy density decreases monotonically as one approaches the outer boundary of the thin shell. Hence, the total mass of the thin shell can be readily evaluated using the expression for the surface energy density, which is given by,
\begin{equation}
    m_s=4\pi R^2 \varsigma
\end{equation}
\begin{widetext}
\begin{equation}
    m_s=-D\left( \sqrt{1-\frac{2M}{D}+\frac{q^2}{D^2}}-\sqrt{1-\frac{\alpha T_0^2 D^2}{6 F}-\frac{ch\pi^3}{90 F D^2}-\frac{8 \pi \rho_0 D ^2}{6 F}-\frac{Q^2}{F D^2} }\right).
\end{equation}    
\end{widetext}
where M is the total mass of the gravastar. 
\section{Physical features of gravastar in this model.}\label{section6}
In this section, we examine the physical properties of the constructed gravastar model, including proper length, entropy, stability, and energy levels within the shell region. Since the geometry of the gravastar is formed by matching two distinct spacetime manifolds, the stiff matter fluid constituting the thin shell evolves consistently across these regions. The dynamical behavior of this fluid is, therefore, governed by the junction conditions imposed at the shell.

\subsection{Proper length}

Following the proposal of Mazur and Mottola\citep{Mazur:2001fv}, the shell of the gravastar is composed of stiff matter and is located between two distinct spacetime regions. This shell occupies the intermediate zone that separates the interior core from the exterior spacetime. Specifically, the inner boundary of the shell is defined at $r=D$, which corresponds to the phase transition surface between the interior region and the thin shell, while the outer boundary is located at $r=D+\epsilon$, where a proper thickness of the shell can be assumed to be small $\epsilon<<1$ representing the transition between the shell and the exterior spacetime. Accordingly, the proper thickness of the shell, as well as the physical separation between these two phase boundaries, can be evaluated using the following expression.
\begin{equation}
    l=\int_{D}^{D+\epsilon} \sqrt{e^\lambda}dr,
\end{equation}
\begin{equation}
     l=\int_{D}^{D+\epsilon}\frac{1}{\sqrt{\frac{-h c \pi^2}{45 r^2}+\frac{Q^2}{r^2}+c_2+ 2 \log r }}dr .
     \label{eq: length}
\end{equation}

\begin{figure}[!]
    \centering 
    \includegraphics[width=01\linewidth]{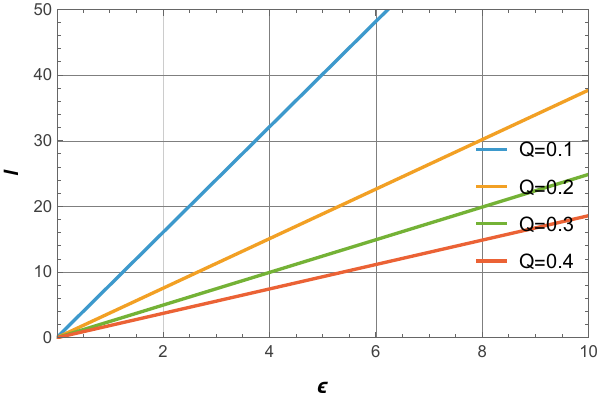}
    \caption{ The variation of proper length $l$ with respect to shell thickness $\epsilon$ for different values of charge $Q$ is illustrated in Fig. 3 }
    \label{fig3}
\end{figure}
\autoref{fig3} displays the behaviour of the proper length '$l$' as a function of the shell thickness parameter $\epsilon$ for different values of the charge $Q$. In all cases $l$ increases monotonically and essentially linearly with $\epsilon$, confirming that the shell geometry is regular and free from any divergence over the range considered. For a fixed value of $\epsilon$, the proper length decreases as the charge $Q$ increases: the curve for $Q=0.1$ lies above those for $Q=0.2,0.3$ and $0.4$ in that order. This behaviour follows directly from  Eq. \eqref{eq: length}, where the electromagnetic term $Q^2/r^2$ enters the integrand with a sign that increases $e^{-\lambda}(r)$, i.e. it strengthens the metric potential $g_{rr}^{-1}$ appearing under the square root, thereby reducing $\sqrt{e^{\lambda}}$ and hence the proper distance corresponding to a given coordinate interval. Physically, the repulsive electromagnetic contribution partially compensates the geometric stretching produced by the Casimir term, so that a more highly charged configuration has a comparatively 'thinner' shell in proper-distance terms for the same coordinate width $\epsilon$. The finiteness and smooth, single-valued nature of $l(\epsilon)$ for every $Q$ considered indicates that the thin-shell approximation remains self-consistent and that no horizon or curvature singularity develops within the shell.

\subsection{Energy Content}
In gravastar models, the thin shell plays a crucial role in balancing the repulsive effects of the interior vacuum region with the gravitational attraction of the exterior spacetime. The energy stored in this shell originates from the surface stresses required to maintain this equilibrium. Therefore, evaluating the total energy of the thin shell is essential for understanding the physical viability and stability of the gravastar configuration.
\begin{equation}
    E_{shell}=\int_{D}^{D+\epsilon} 4 \pi r^2 \rho dr,
\end{equation}

\begin{widetext}
\begin{equation}
\begin{aligned}
E_{shell} = \int_{D}^{D+\epsilon} 4\pi r^2 \Bigg[ & e^{-r\nu} c_1
+ \frac{1}{270}\, e^{-r\nu} \Bigg( \frac{1}{r^4}\, e^{r\nu}
\Big( 540 r^2 (1+3r\nu) + ch\,\pi^2 \left(6+10 r\nu+5 r^2\nu^2+5 r^3\nu^3\right) \\
& \quad - 45 Q^2 \left(6+10 r\nu+5 r^2\nu^2+5 r^3\nu^3\right) \Big) \Bigg)   -5\nu^2 \left(324 + ch\,\pi^2\nu^2 - 45 Q^2\nu^2\right) \operatorname{Ei}(r\nu)  \Bigg] dr
\end{aligned}
\end{equation}
\end{widetext}

\begin{figure}[!ht]
    \centering 
    \subfloat[]{\includegraphics[width=1\linewidth]{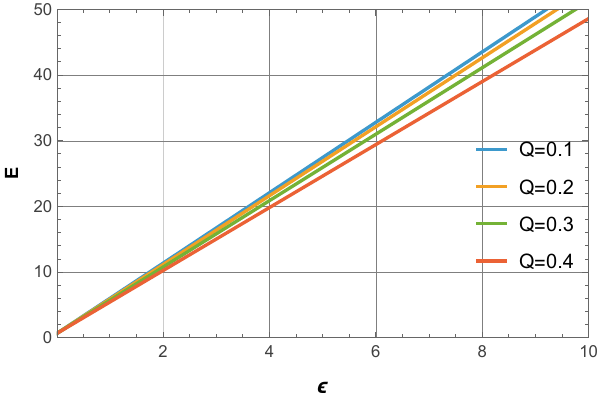}}
    \caption{{Variation of the energy $E$ inside the thin shell  with respect to the thickness of the thin shell $\epsilon$ for different values of charge Q is illustraed in Fig. 4} }
    \label{fig4}
\end{figure}

\autoref{fig4} shows that, the energy increases monotonically and smoothly with $\epsilon$ for every $Q$, with no sign of divergence or discontinuity, indicating that the matter-energy distribution within the shell is well-behaved throughout its extent. As in the case of the proper length, larger values of $Q$ correspond to slightly lower energy content for a given $\epsilon$, since the electromagnetic contribution enters the density expression, Eq. \eqref{eq: p=r}, with a negative coefficient in the dominant terms, partially offsetting the positive contributions from the Casimir and stiff-matter sectors. However, the separation between the curves is markedly smaller than in \autoref{fig3}, showing that the total energy of the shell is far less sensitive to the charge parameter than the proper length is; the charge principally reshapes the local geometry (via $e^{-\lambda}$) rather than the density profile $\rho(r)$ itself. The monotonic, near-linear growth of $E_{shell}$ with $\epsilon$ is consistent with a physically reasonable, positive-energy shell and supports the structural stability of the gravastar configuration.
\subsection{Entropy}
The entropy associated with the thin shell provides important insight into the thermodynamic behavior and microscopic structure of the gravastar. Since the shell is composed of ultra-relativistic (stiff) matter and separates two distinct spacetime regions, its entropy plays a key role in characterizing the statistical degrees of freedom localized on the junction surface. Unlike black holes, gravastars do not possess an event horizon; therefore, the entropy arises entirely from the matter content of the shell rather than from horizon area.
\begin{equation}
    S=\int_{D}^{D+\epsilon}4\pi r^2 s(r) \sqrt{e^\lambda} dr,
\end{equation}
where
\begin{equation}
    s(r)=\alpha \sqrt{\frac{p}{2\pi}}.
\end{equation}
\begin{figure}[!ht]
    \centering 
    \includegraphics[width=1\linewidth]{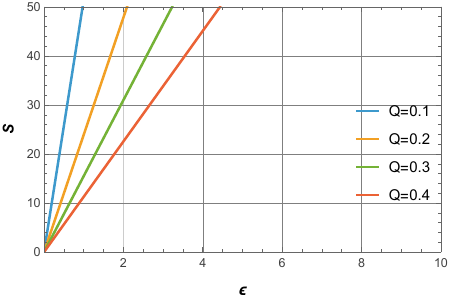}
    \caption{ {Variation of the entropy $S(r)$ within the gravastar shell as a function of the radial coordinate $r$. The smooth increase indicates a finite and well-defined entropy distribution across the shell region.} }
   \label{fig5} 
\end{figure}
\autoref{fig5} illustrates the variation of the entropy $S$ of the thin shell with the shell thickness $\epsilon$ for different values of the charge parameter $Q$. As with the proper length and energy, the entropy grows monotonically and smoothly with $\epsilon$ for every value of 
$Q$, confirming a finite, well-defined entropy density throughout the shell with no divergence at either boundary.Entropy $S$ rises smoothly and monotonically with $\epsilon$ for all $Q$ , but the growth rate depends strongly on charge: $Q=0.1$ saturates the plotted range by 
$\epsilon \approx 2$ , while $Q=0.4$ needs $\epsilon \approx 5$ to reach the same value. This follows because entropy density is weighted by 
$\sqrt{e^{\lambda}}$, which higher $Q$ suppresses, so more charged shells need greater coordinate thickness to build up the same entropy. 

\section{Conclusion}
In this work, we have constructed and investigated a charged gravastar configuration within the framework of modified teleparallel gravity, considering the quadratic form $f(T)= T +\alpha T^2$ and incorporating Casimir vacuum energyinto the interior region. The gravastar is described by three distinct regions: a Casimir energy dominated interior, an intermediate thin shell of ultra-relativistic matter satisfying the equation of state $(p=\rho)$, and an exterior Reissner Nordstrom spacetime. The inclusion of the Casimir contribution provides a negative vacuum energy density, while the electromagnetic field introduces pressure anisotropy and modifies the gravitational structure of the configuration.
\\
For the interior region, the equation of state $(p=-\rho)$ leads to a constant torsion scalar for the nonlinear $(f(T))$ model. The resulting interior metric contains distinct contributions from the matter density, the quadratic torsion correction, the Casimir energy, and the electric charge. The regularity condition $(C_{1}=0)$ eliminates the $(1/r)$ singular term at the origin, thereby yielding a regular interior geometry. In the limit $(\alpha\rightarrow0)$, the solution consistently approaches the corresponding teleparallel/GR behaviour.
\\
The intermediate region is modeled as a thin shell of stiff matter. Under the thin-shell approximation, the metric functions and the corresponding energy density are obtained, and the resulting density and pressure profiles remain positive and smoothly decrease with increasing radial coordinate. This indicates a well-behaved matter distribution within the shell. The exterior region is described by the Reissner-Nordstrom geometry, allowing the effects of the electric charge to be incorporated consistently into the complete gravastar spacetime.
\\
The matching between the interior and exterior geometries has been carried out using the Israel-Darmois junction formalism. The resulting surface energy density and pressure characterize the stresses required to maintain the thin shell and provide a consistent transition between the two spacetime regions. The corresponding energy conditions were also considered to examine the physical acceptability of the shell.
\\
Furthermore, we have examined several physical properties of the shell, including its proper length, energy content, and entropy. The proper length remains finite and increases smoothly with the shell thickness, while increasing charge reduces the proper length for a fixed coordinate thickness. Similarly, the shell energy increases smoothly with its thickness and remains finite over the region considered. The entropy also exhibits a finite and monotonic behaviour, with its dependence on the charge parameter reflecting the modification of the shell geometry by the electromagnetic field. These results indicate that the shell remains regular and physically well behaved throughout the parameter range investigated.
\\
Overall, the combined effects of Casimir vacuum energy, electric charge, and nonlinear torsion corrections provide a consistent framework for constructing a regular compact-object configuration without the central singularity and event horizon associated with the classical black-hole picture. The finite proper length, energy, and entropy of the shell, together with the stability analyses considered in this work, support the physical viability of the proposed charged Casimir-supported gravastar. The model therefore demonstrates that quantum vacuum effects coupled with modified teleparallel gravity can play an important role in the structure and stability of ultra-compact objects.

\section*{Acknowledgments}

CSV and VV acknowledge DST, New Delhi, India, for supporting research facilities under DST-FIST-2019.

\section*{Data Availability}

No new data were generated or analyzed in this study.

\bibliography{Ref.bib}

\end{document}